\documentstyle[eqsecnum,aps]{revtex}

\begin{document}
\title{{\bf The evolution and revival structure of angular momentum quantum wave
packets }\\
(Tutorial)}
\author{Marcis Auzinsh}
\address{Department of Physics,\\
University of Latvia, Rainis blvd. 19, \\
Riga LV--1586, Latvia}
\maketitle

\begin{abstract}
In this paper a coherent superposition of angular momentum states created by
absorption of polarized light by molecules is analyzed. Attention is paid to
the time evolution of wave packets representing spatial orientation of
internuclear axis of diatomic molecule. Two examples are considered in
detail. Molecules absorbing light in a permanent magnetic field experiencing
Zeeman effect and molecules absorbing light in a permanent electric field
experiencing quadratic Stark effect. In a magnetic field we have a wave
packet that evolves in time exactly as classical dipole oscillator in a
permanent magnetic field. In the second case we have the wave packet that
goes through periodical changes of a shape of the packet and revivals of
initial shape. This is a pure quantum behavior. Classical motion of angular
momentum in an electric field in case of quadratic Stark effect is known to
be aperiodic. Obtained solutions for wave packet evolution are briefly
compared with Rydberg state coherent wave packets and harmonic oscillator
wave packets.
\end{abstract}

\section{Introduction}

There are some problems that can be found in every classical mechanics text
book. For example, rotation of the planets around the Sun under the action
of the gravitation force or the oscillations of the pendulum under the
action of the quasielastic force.

In quantum mechanics there are very similar problems of the same importance.
The motion of the electron around the nucleus under the action of Coulomb
force or the vibration of a diatomic molecule along the line connecting both
nuclei. To compare these two sets of problems from which the first belongs
to the macroscopic world and the other to the microscopic world one can ask
the questions of a type: Is it possible to observe the motion of an electron
in a Kepler orbit around the nucleus in the same way as it is possible to
observe the motion of the planet around the Sun? That is, is it possible to
obtain experimentally a Rutherford atom when electron rotates around the
nucleus in a Kepler orbit? Is it possible to observe oscillations of the
nuclei in a molecule that are similar to classical oscillations of point
particle bound by a quasielastic force? An affirmative answer to these
questions, as is well known, is given by the correspondence principle of
quantum mechanics \cite{av1}. \ 

The most common way how we can experimentally examine the objects in the
micro world is by their interaction with light. At the same time it is
common wisdom, see for example \cite{av1}, that the methods of ordinary
optical spectroscopy generally involve excitation of individual stationary
states of atoms and molecules. Such states describe objects that are quantum
mechanical by nature. For example, even for arbitrarily large quantum
numbers a single stationary state of an electron in a Coulomb field does not
come close to describing the motion of a localized particle in a Kepler
orbit, just as for any quantum number a stationary wave function of a
harmonic oscillator does not describe the harmonic oscillations of a
localized particle. In fact, classical motion is never obtained from
excitation of single quantum state.

Recently, with the use of ultrashort optical pulses, it has become possible
to create coherent superpositions of many quantum states to obtain localized
wave packets that are particle-like objects that obey quasiclassical laws,
see \cite{al1,ga1} and references cited therein. Usually electrons moving in
orbits with large radiuses and oscilations of molecules are examined in
these experiments. The processes with these wave packets usually are very
fast -- often occurring on a pico- or even on a femto-second time scale.
These types of processes are most often analyzed to examine correspondence
between the classical and quantum description of the objects in the
microscopic world.

The interaction of particles with definite angular momentum with electric
and magnetic fields provides a second, more accessible, but less exploited,
way to examine the correspondence between the quantum and classical nature
of the microscopic world. For example, from the view point of classical
physics the angular momentum vector of a charged particle spinning in a
magnetic field will precess around the field direction with the Larmor
frequency \cite{au4} 
\begin{equation}
\omega _L=\frac{g_J\mu _BB}\hbar ,  \label{eq0}
\end{equation}
preserving the projection of the angular momentum on the direction of the
field. Here $g_J$ is a Lande factor, $\mu _B$ is the Bohr magneton, and $B$
is the magnetic field strength. To determine the behavior of a particle --
atom or molecule -- in an external magnetic field in quantum mechanics one
must deal with the particle's angular momentum states and corresponding wave
functions $Y_{JM}(\theta ,\varphi )$. To obtain classical-like motion of
quantum angular momentum in an external field one must analyze superposition
of angular momentum states. For diatomic molecules squared modulus of this
superpositional wave packet $\Psi \left( \theta ,\varphi ,t\right) $ will
show the probability to find molecular axis with a certain orientation in
space. Namely $\left| \Psi \left( \theta ,\varphi ,t\right) \right| ^2\sin
\theta d\theta d\varphi $ is a probability to find molecular axis in the
direction in space characterized by spherical angles $\theta $ and $\varphi $%
. For the case of molecules this approach has another advantage. To compare
classical and quantum results usually one wants to examine the behavior of
the system as the angular momentum become large. According to the
correspondence principle these states should behave classically. For the
case of rotational states of molecules it is very common and straightforward
to create states with large angular momentum quantum numbers, with $J\propto
100$ typical \cite{au1}. The situation is different for the case of atomic
Rydberg states. It is not at all easy to create in a laboratory states with
principal quantum number $n$ close to $100$ \cite{al1}.

Partially the problem to obtain coherent superposition of angular momentum
states was solved already many years ago, when quantum beats in an external
magnetic field were observed experimentally for the first time by
Aleksandrov in Russia \cite{alx1} and by Dodd with coworkers in United
Kingdom \cite{do1}. They used light pulse to excite coherently several
angular momentum states $\left| J,M\right\rangle =Y_{JM}\left( \theta
,\varphi \right) $ with the same angular momentum quantum number $J$, but
different magnetic quantum numbers $M$ simultaneously and coherently. In an
external magnetic field these angular momentum states have different
energies $E_M$ and hence the corresponding wave functions have different
phase factors $\exp [-i(E_M$/$\hbar )t]$. In an experiment one can observe
harmonic time dependencies of polarized fluorescence that corresponds to the
beats between these wave functions with different phase factors in the same
way as in a signal processing one can observe beets between two or more
harmonic signals with slightly different frequencies.

From the practical view point these experiments can be more straightforward
than experiments with Rydberg state atoms or instant excitation of
oscillations in molecules. An obvious reason for this is that speed of the
processes in an external field usually are slower and in any case it can be
controlled by controlling external field strength. As a result, for
excitation of the state one can use much longer laser pulses and also
observation of the state dynamics can be performed by much slower
experimental devices \cite{au1}.

In this paper quantum beat experiments will be analyzed by considering the
creation and time evolution of angularly localized wave packets. This unique
approach provides an opportunity to examine the correspondence between
classical and quantum--mechanical periodic motion as induced by electric and
magnetic fields. In our knowledge from these positions these experiments
have not been analyzed before.

\section{Revival structure of wave packets}

The time-dependant wave functions for angularly localized angular momentum
quantum wave packets formed as a coherent superposition of angular momentum
eigenstates may be written as 
\begin{equation}
\Psi \left( \theta ,\varphi ,t\right) =\sum_Mc_MY_{JM}\left( \theta ,\varphi
\right) \exp \left( -i\frac{E_M}\hbar t\right) ,  \label{eq1}
\end{equation}
where $Y_{JM}\left( \theta ,\varphi \right) $ is an ordinary spherical
function \cite{va1} and the coefficients $c_M$ are complex amplitudes.

We would like to examine to what extent and for how long time evolution of
this wave packet coincides with the predictions of classical physics. For
example, classically, an angular momentum of a rotating rigid charge
distribution will precess in the magnetic field with the Larmor frequency $%
\omega _L$ $\left( \ref{eq0}\right) $. If there is a coincidence between the
quantum and classical discription, then the position of the ``center of
gravity'' of a spatially localized wave packet (the average value of the
particle angular coordinates) must precess in space according to the rules
of classical mechanics \cite{au4} 
\begin{equation}
\frac{d{\bf J}}{dt}={\bf \mu \times B.}  \label{eq1a}
\end{equation}
Here ${\bf J}$ is a classical angular momentum vector, ${\bf \mu }$ ---
magnetic dipole moment of particle, ${\bf B}$ --- strength of the external
magnetic field.

Because classical particles have a definite direction of angular momentum,
wave packets localized in space usually have $c_M$ well centered around some
particular mean quantum number $\overline{M}.$ For a similar reason wave
packets that can be created from stationary atomic Rydberg state wave
functions are centered around some definite value $\overline{n}$ of the
principle quantum number $n$ of the atomic state. Rydberg wave packets
weighted by coefficients possessing Gaussian distribution \cite{bl1} are
particularly well investigated: 
\begin{equation}
\left| c_n\right| ^2=\frac 1{\sqrt{2\pi }\sigma }e^{-(n-\overline{n}%
)/2\sigma ^2}.  \label{eq4}
\end{equation}
Here the parameter $\sigma $ characterizes the width of this distribution.
Another particularly well investigated case is the coherent states of a
harmonic oscillator that can be obtained from harmonic oscillator wave
functions $\left| v\right\rangle $ weighted by coefficients in the form \cite
{bl1} 
\begin{equation}
c_v=e^{-(1/2)\left| \alpha \right| ^2}\frac{\alpha ^v}{\sqrt{v!}},
\label{eq5}
\end{equation}
where $v$ is the vibration quantum number and $\alpha $ is a parameter.

What we are interested in most, when we think about evolution of wave
packets is, what are the laws governing the long-term postclassical
evolution of wave packets beyond the bounds of the dynamics according to the
correspondence principle?

The assumption that the weighting probabilities $\left| c_M\right| ^2$ are
strongly centered around a mean value $\overline{M}$ (or $\overline{n},%
\overline{v}$) means that only those states with energies $E_M$ near the
value $E\overline{_M}$ enter appreciably into the sum of Eq. $\left( \ref
{eq1}\right) $. This permits an expansion of energy in a Taylor series in $M$
around the centrally excited value $\overline{M}$%
\begin{equation}
E_M=E\overline{_M}+E^{\prime }\overline{_M}(M-\overline{M})+\frac 12%
E^{\prime \prime }\overline{_M}(M-\overline{M})^2+\frac 16E^{\prime \prime
\prime }\overline{_M}(M-\overline{M})^3+...,  \label{eq6}
\end{equation}
where each prime on $E\overline{_M}$ denotes a derivative at point $M=%
\overline{M}$.

The derivative terms in Eq. $\left( \ref{eq6}\right) $ define distinct time
scales \cite{av1,bl1} 
\begin{equation}
T_{cl}=\frac{2\pi }{\left| E^{\prime }\overline{_M}\right| },\qquad t_{rev}=%
\frac{2\pi }{\frac 12\left| E^{\prime \prime }\overline{_M}\right| },\qquad
t_{sr}=\frac{2\pi }{\frac 16\left| E^{\prime \prime \prime }\overline{_M}%
\right| }.  \label{eq7}
\end{equation}
The first time scale $T_{cl}$ is called the classical period. It is the
period after which system returns to it's initial position according to the
laws of classical physics. The second time scale $t_{rev}$ is the revival
time. This is a time after which the initial wave function will be partially
or completely rebuilt. The third time scale $t_{sr}$ is the superrevival
time. It represents the time after which the wave function will be rebuilt
in case that it is only partially rebuilt after the revival time. For the
most commonly investigated situations of the coherent superposition of
Rydberg states and anharmonic oscillators the timescales are ordered so that 
$T_{cl}\ll t_{rev}\ll t_{sr}$. But, as we will see, this time ordering
changes for the coherent superposition of angular momentum states in an
electric field.

One particularly good thing about evolution of angular momentum wave packets
is that we can know {\em exactly} and to some extent, control by changing
excitation geometry and polarization of excitation light, the $c_M$
distribution that will occur in a realistic experiment.

As an example let us consider a so called $Q$-type of molecular transition
when light excites molecules from the ground state to the excited state and
both states have the same value of angular momentum quantum number $J$. Let
us further assume that the exciting radiation is linearly polarized with the
light vector ${\bf e}$ lying perpendicularly to an external magnetic field $%
{\bf B}$. The probability of finding molecules in a particular angular
momentum state $Y_{JM}\left( \theta ,\varphi \right) $ can be found by
determining the diagonal elements of the density matrix $f_{MM}$ which give
the population of angular momentum substates characterized by a magnetic
quantum number $M$ (for details see Appendix I and \cite{au4,au1}) 
\begin{eqnarray}
f_{MM} &=&\left| c_M\right| ^2=\sum_{{\bf \mu }}\left| \left\langle M\right| 
{\bf e}^{*}{\bf d}\left| \mu \right\rangle \right| ^2=  \nonumber \\
&=&\frac 12\left[ \left( C_{11JM-1}^{JM}\right) ^2+\left(
C_{1-1JM+1}^{JM}\right) ^2\right] =\frac 12-\frac{M^2}{2J(J+1)}.  \label{eq2}
\end{eqnarray}
Here $\left\langle M\right| {\bf e}^{*}{\bf d}\left| \mu \right\rangle $ is
the optical transition matrix element, ${\bf d}$ is the optical transition
dipole moment operator, $C_{11JM-1}^{JM}$ is the Clebsch--Gordan coefficient 
\cite{va1}, and $\mu $ is a ground state magnetic quantum number. In this
expression Clebsch--Gordan coefficients of a type $C_{11JM-1}^{JM}$
represent quantum mechanical amplitude to excite an angular momentum state $%
\left| J^{\prime },M\right\rangle $ from an initial (usually ground) state $%
\left| J^{\prime \prime },\mu =M-1\right\rangle $. In this particular case
of a $Q$-type molecular transition the absorption of light does not change
the angular momentum of the molecule or atom, so $J^{\prime \prime
}=J^{\prime }=J$. As an example for $J=20$, $f_{MM}=\left| c_M\right| ^2$ as
calculated from Eq. $\left( \ref{eq2}\right) $ is given in Figure. 1

Off diagonal elements of the density matrix represent the coherence (phase
relations) between different angular momenta substates. Off diagonal matrix
elements can be calculated as 
\begin{equation}
f_{MM^{\prime }}=\overline{c_Mc_{M^{\prime }}^{*}}=\sum_{{\bf \mu }%
}\left\langle M\right| {\bf e}^{*}{\bf d}\left| \mu \right\rangle
\left\langle M^{\prime }\right| {\bf e}^{*}d\left| \mu \right\rangle ^{*}
\label{eq7a}
\end{equation}
For $Q$-excitation with a pulsed light polarized along $y$ axis besides the
diagonal matrix elements calculated according to the Eq. $\left( \ref{eq2}%
\right) $ we will have the following non-zero offdiagonal matrix elements 
\cite{au4} 
\begin{eqnarray}
f_{M+1,M-1} &=&f_{M-1,M+1}=\frac 12C_{JM1-1}^{JM-1}C_{JM11}^{JM+1}  \nonumber
\\
&=&-\frac{\sqrt{\left( J^2-M^2\right) \left[ \left( J+1\right) ^2-M^2\right] 
}}{4J(J+1)}.  \label{eq7b}
\end{eqnarray}
From the density matrix elements for angular momentum states, we can easily
calculate the squared wave function that represents the probability density 
\begin{equation}
\left| \Psi \left( \theta ,\varphi ,t\right) \right| ^2=\frac 3{2J+1}%
\sum_{MM^{^{\prime }}}f_{MM^{\prime }}Y_{JM}Y_{JM^{\prime }}^{*}\exp
(-i\omega _{MM^{^{\prime }}}t),  \label{eq7c}
\end{equation}
where $\omega _{MM^{^{\prime }}}=\left( E_M-E_{M^{^{\prime }}}\right) /\hbar
.$

\section{Angular momentum wave packets in a magnetic or an electric field}

\subsection{Atom or molecule in the external magnetic field}

If an atom or molecule is in an external magnetic field experiencing
ordinary Zeeman effect the angular momentum state's $J$ magnetic sublevels $%
\left| J,M\right\rangle $ with different magnetic quantum numbers $M$ will
have energies 
\begin{equation}
E_M(M)=E^{(0)}+g_J\mu _BBM/\hbar =E^{(0)}+E_LM.  \label{eq8}
\end{equation}
Where $E^{(0)}$ is the energy of a state in the absence of the external
field. According to Eq. $\left( \ref{eq7}\right) $ we can expect classical
period $T_{cl}$ to be equal to $2\pi /E_L=2\pi /(g_J\mu _BB/\hbar )$. All
other periods will be infinite. Thus time evolution of this angular momentum
wave packet will be infinitely long rotation around the external magnetic
field direction with Larmor angular frequency $\omega _L=2\pi /T_{cl}$. No
changes apart from the rotation in space around the magnetic field direction
will occur to the wave function. This is what actually has been observed
experimentally in the past \cite{qb1} as a harmonic modulation of an
intensity of a polarized fluorescence from an ensemble of atoms or molecules
excited by a short laser pulse. This is a well known effect of quantum beats
induced by magnetic field.

From the view point of an evolution of a wave packet, this result is similar
to the well known behavior for coherent states of a harmonic oscillator. In
his pioneering paper Ervin Schrodinger wrote as early as in 1926, that wave
packets formed as coherent states of harmonic oscillator will oscillate
infinitely long between classical turning points without dispersion, for
references see \cite{bl1}. The main reason for this is that the energies of
quantum harmonic oscillator states depend{\em \ linearly }on the vibration
quantum number $v$. It means that only the first derivative $E^{\prime }%
\overline{_v}$ in the expansion of type $\left( \ref{eq6}\right) $ will
differ from zero. In case of Zeeman effect we observe the same linear energy
dependence of the magnetic sublevels of atomic or molecular states on the
magnetic quantum number $M$. As a result, according to Eq. $\left( \ref{eq7}%
\right) $ we have infinitely long classical type motion of the wave packet
that represent the precession of angular momentum in a magnetic field. One
can easily calculate angular momentum distribution after the pulsed
excitation following equations $\left( \ref{eq7a}\right) $ -- $\left( \ref
{eq7c}\right) $. This result appears to be independent of the value of
angular momentum quantum number $J$. It is 
\begin{equation}
\left| \Psi \left( \theta ,\varphi ,t\right) \right| ^2=\frac 3{8\pi }\left[
1-\sin ^2\theta \sin ^2(\varphi -\omega _Lt)\right] .  \label{eq8a}
\end{equation}
We have donut shape wave function that rotates in space around $z$ axis with
Larmor frequency $\omega _L$, Eq. $\left( \ref{eq0}\right) $ \cite{au1}, see
Fig. 2. The fact that the result is independent of the angular momentum
quantum number $J$ is worth to mention. It is interesting especially because
this distribution coincides precisely with the result that would appear if
we would consider absorption of electric dipole oscillator in classical
physics in the same circumstances. Indeed, if instead of considering
artificial wave packets that can only be studied theoretically, we consider
wave packets that can be obtained in a realistic experiment it is not
uncommon for the quantum and classical results coincide even for small
quantum numbers, see \cite{au1}. In classical approach this donut shape
disrtibution of molecular axis can be understood if one keeps in mind that
for $Q$-type molecular transition in a classical approach absorbing dipole
is oriented along ${\bf J}$, it means perpendicularly to the intermolecular
axis of a rotationg molecule \cite{au1}.

\subsection{Molecule in an external electric field}

The evolution of a molecular wave packet in an external electric field is
quite distinct from the case of a magnetic field. For both cases we have the
same amplitudes $c_M$ of partial components of wave function in Eq. $\left( 
\ref{eq1}\right) $ and the same density matrix. But in case of an electric
field we will have a different magnetic sublevel energy $E_M$ dependence on
the magnetic quantum number $M$. Let us consider molecule in a state
experiencing a quadratic Stark shift. This is a type of Stark effect most
commonly observed with atoms and molecules. In this case we will have an
energy dependence on the magnetic quantum number of the form 
\begin{eqnarray}
E_M(M) &=&E^{(0)}+\frac{d^2{\cal E}^2}{hB}\left[ \frac{J(J+1)-3M^2}{%
2J(J+1)(2J-1)(2J+3)}\right]  \nonumber \\
&=&E^{(0)}+E_{Stark}^{(1)}({\cal E)}+E_{Stark}^{(2)}({\cal E)}M^2.
\label{eq9}
\end{eqnarray}
We know that in case of the Stark effect, the classical motion of angular
momentum in an external electric field is aperiodic \cite{au4,hi1}. This is
exactly what we can see from Eq. $\left( \ref{eq9}\right) $. The first
derivative of $E_{\overline{M}=0}^{\prime }$ is zero and that means that $%
T_{cl}$ is infinite. At the same time the second derivative 
\begin{equation}
E_{\overline{M}}^{\prime \prime }=-\left[ \frac{3d^2{\cal E}^2}{%
hBJ(J+1)(2J-1)(2J+3)}\right]  \label{eq10}
\end{equation}
differs from zero and one can expect $t_{rev}$ be different from zero or
infinity. This is exactly what was predicted for Stark quantum beets \cite
{au2}.\ Figure 3 depicts one period of evolution of the wave function for
state $J=1$ excited by linearly polarized light with ${\bf e}$ vector lying
in $zy$ plane and forming angle $\pi /4$ with the direction of an external
electric field ${\cal E}$. The $R$-type $\left( J=0\longrightarrow
J=1\right) $ optical transition is assumed. The analytical expression
describing probability density on Figure 3 is 
\begin{equation}
\left| \Psi (\theta ,\varphi ,t)\right| ^2=\frac 3{8\pi }\left\{ 1-\sin
^2\theta \cos ^2\varphi +\sin 2\theta \sin 2\varphi \cos [(2\pi
/t_{rev})t]\right\} .  \label{eq10a}
\end{equation}
It can be calculated using formulae $\left( \ref{eq7a}\right) $, $\left( \ref
{eq7c}\right) $ and Appendix.

The interesting feature of Stark effect is that revival time 
\begin{equation}
t_{rev}=\frac{4\pi hBJ(J+1)(2J-1)(2J+3)}{3d^2{\cal E}^2}  \label{eq11}
\end{equation}
approaches infinity when angular momentum approaches infinity, it means a
particle with very large angular momentum starts to behave truly
classically. This quantum mechanical revival is not only interesting as a
peculiar behavior of wave function, but as well can be used to orient
molecules in beams effectively \cite{au3}. Probably other applications of
periodic behavior of atomic and molecular wave function in electric field
can be foreseen.

\section{Summary}

This paper points attention and illustrates two examples of angular momentum
coherent wave packets that are of grate interest due to their peculiar
properties. They are less noticed than they deserve. These wave packets
describe precession of internuclear axes of diatomic molecules in an
external field.

The first example was the angular momentum coherent superposition state
created by the absorption of polarized light by a molecule in an external
magnetic field. There are many examples of such wave packets created in
experiments for states of diatomic molecules \cite{qb1}. But never has this
situation of a coherent superposition of angular momentum eigenstates been
analyzed with the same machinery used to analyze the coherent superposition
of Rydbaerg states or coherent states of harmonic oscillator. In experiments
with molecules absorbing light in permanent external fields very often
states with large rotational angular momentum quantum numbers of $J\propto
100$ were involved. This allows us to compare these states with behavior of
a spinning particle in an external field.

It is known that classical angular momentum in an external magnetic field
will precess around the magnetic field direction with Larmor frequency $%
\omega _L$.

Quantum wave packet in a magnetic field will experience the same motion.
Period of rotation of wave packet will coincide perfectly with period of
precession of classical angular momentum. Wave packet will last for ever
(actually as long as the excited state of molecule will live). It will not
undergo any disintegration.

There is known only one other example when the wave packet evolves in time
without dispersion. It is a coherent state of harmonic oscillator \cite{bl1}%
. The reason for this type of motion in both cases is the same. All
coherently excited wave functions in these examples represent states that
are equally separated in energy scale, i.e. the systems have energy levels
with equally separated steps.

Another example considered here was angular momentum states in an external
electric field causing a quadratic Stark effect. In this case an ensemble of
angular momentum will evolve aperiodically in classical physics \cite{au4}.
In quantum physics we will have periodical motion during which the wave
function will periodically disintegrate and than, after a definite period,
will go through a revival.

This is quite unique dynamics. It is more usual for systems to have a period
of classical motion $T_{cl}$ that is substantially shorter than revival time 
$t_{rev}$, as is true for the Rydberg states or anharmonic oscillator
states. It means that during one revival period this type quantum system
will undergo many classical periods. In the example of the quadratic Stark
effect in an external electric field, we have exactly the opposite extreme.
The system has no classical period at all. From the view point of classical
physics the system is aperiodic. But quantum evolution of the wave function
still have a well defined period which becomes longer and longer, when
angular moment of system increases and system approaches classical limit.

\section{Appendix I. Calculation of density matrix elements.}

Calculation of density matrix elements entering Eqs. $\left( \ref{eq2}%
\right) $ and $\left( \ref{eq7a}\right) $

\begin{equation}
f_{MM^{\prime }}=\sum_{{\bf \mu }}\left\langle M\right| {\bf e}^{*}{\bf d}%
\left| \mu \right\rangle \left\langle M^{\prime }\right| {\bf e}^{*}d\left|
\mu \right\rangle ^{*}  \label{eqa1}
\end{equation}
mainly consist in calculation of quantum mechanical matrix elements of a type

\begin{equation}
\left\langle M\right| {\bf e}^{*}{\bf d}\left| \mu \right\rangle .
\label{eqa2}
\end{equation}
Let us now have a look in more detail at how these matrix elements can be
calculated. The first step is to calculate the {\em Hermitian product} \cite
{fa1} $\left( {\bf e}^{*}\cdot {\bf d}\right) $ of light polarization vector 
${\bf e}$ and optical transition dipole moment ${\bf d}$. Meaning of this
product is very close to that of a scalar or dot product of two ordinary
vectors. Only in this case we are dealing with complex vectors and to find
the projection of one vector onto another or -- which is the same -- to find
``how much of one vector is contained into another'' we must calculate a
Hermitian product.

We are using here complex vectors, because it is a simple way to describe
rotations in quantum mechanics as well as in classical physics. Let us see
how these complex vectors and their components in a {\em cyclic system of
coordinates} \cite{au1,va1} can be used to describe, for example, the
polarization of light. An arbitrary light polarization vector (unit vector $%
{\bf e}$ in the direction of the electric field vector of the light wave) in
cyclic coordinates can be written as

\begin{eqnarray}
e^{+1} &=&-\frac 1{\sqrt{2}}(e_x-ie_y),  \nonumber \\
e^0 &=&e_z,  \label{eqa3} \\
e^{-1} &=&\frac 1{\sqrt{2}}(e_x+ie_y).  \nonumber
\end{eqnarray}
If we multiply now these vector components by a phase factor $\exp \left(
-i\Omega t\right) $ that represents the oscillations of electric field in
light wave, we can easily see, that for 
\begin{eqnarray}
e^{+1}\exp \left( -i\Omega t\right) &=&-1/\sqrt{2}[e_x\exp \left( -i\Omega
t\right) -ie_y\exp \left( -i\Omega t\right) ]  \nonumber \\
&=&-1/\sqrt{2}\{e_x\exp \left( -i\Omega t\right) +e_y\exp [-i(\Omega t+\pi
/2)]\}  \label{eqa3a}
\end{eqnarray}
oscillations along $x$ axis are quarter period ahead of those along $y$ axis
and this means that this component of the ${\bf e}$ vector represents
lefthanded circular polarized in $xy$ plane light beam that propagates in
the positive direction of $z$ axis. It means that a light wave in which
electric field vector rotates counterclockwise is viewed in such way that
the radiation approaches the observer in the positive direction of the $z$
axis.

In a similar way one can see that the $e^{-1}$ component represents
righthanded circular polarized light that propagates along $z$ axis with the 
${\bf e}$ vector rotating in a clockwise direction in the $xy$ plane.

And finally, the $e^0$ component represents linearly along $z$ axis
polarized light for which ${\bf e}$ oscillates along $z$ axis. Just as light 
${\bf e}$ vector can be represent by its three Cartesian components, every
vector can be represented by its three cyclic components. In case of light
polarization vector (and many other vectors in quantum mechanics) these
cyclic components are more practical for calculation than the Cartesian
representation.

Now let us turn back to the matrix elements of a type $\left( \ref{eqa2}%
\right) $. According to vector algebra for cyclic vectors, the Hermitian
product can be written as \cite{au1,va1} 
\begin{equation}
\left\langle M\right| {\bf e}^{*}{\bf d}\left| \mu \right\rangle
=\sum_q\left( e^q\right) ^{*}\left\langle M\right| d^q\left| \mu
\right\rangle .  \label{eqa4}
\end{equation}
Further for a matrix element $\left\langle M\right| d^q\left| \mu
\right\rangle $ we can apply Wigner--Eckart theorem \cite{au1,va1}. It
allows to separate angular and dynamical part of this matrix element. What
does this mean? For example, in classical physics if one wants to know how
effectively an oscillating electric field ${\bf e}$ can excite linear dipole
oscillator ${\bf d}$ one must calculate scalar product $\left( {\bf e}\cdot 
{\bf d}\right) =ed\cos \left( \widehat{ed}\right) $. It means that we can
separate dynamic part $ed$ that describes the vector magnitude, and the
angular part $\cos \left( \widehat{ed}\right) $ that represent their mutual
orientation. In the same way in quantum physics Wigner--Eckart theorem
allows us to achieve the same separation for optical transition from initial
state $\left| J^{^{\prime \prime }}\mu \right\rangle $ to excited state $%
\left| J^{\prime }M\right\rangle $. Namely 
\begin{equation}
\left\langle M\right| d^q\left| \mu \right\rangle =\frac 1{\sqrt{2J^{\prime
}+1}}C_{J^{^{\prime \prime }}\mu 1q}^{J^{\prime }M}\left\langle J^{\prime
}\right\| d\left\| J^{^{\prime \prime }}\right\rangle ,  \label{eqa5}
\end{equation}
where $C_{J^{^{\prime \prime }}\mu 1q}^{J^{\prime }M}$ is a Clebsch--Gordan
coefficient and $\left\langle J^{\prime }\right\| d\left\| J^{^{\prime
\prime }}\right\rangle $ is a reduced matrix element, that represents the
dynamical part of optical transition probability. It is obvious that the
Clebsch--Gordan coefficient has a numerical value that depends on the values
of angular momentum projections $M$ and $\mu $ on the quantization axis $z$,
and therefore it describes the angular momentum orientation in space. This
is the angular part of the transition probability.

If we now collect all these formulae together we can obtain the final
expression that allows us to calculate easily matrix elements $\left( \ref
{eqa1}\right) $ for arbitrary light polarization and optical transition
between arbitrary states 
\begin{equation}
f_{MM^{\prime }}=N\sum_{\mu q_1q_2}\left( e^{q_1}\right) ^{*}\left(
e^{q_2}\right) C_{J^{^{\prime \prime }}\mu 1q_1}^{J^{\prime
}M}C_{J^{^{\prime \prime }}\mu 1q_2}^{J^{\prime }M^{\prime }}.  \label{eqa6}
\end{equation}
In this last expression a proportionality coefficient $N$ that is
insignificant for the purpose of this paper, is not determined.

\section{Acknowledgments}

The support from National Research Council Twinning Program, Grant No.
NP--NRC--6224 is greatly acknowledged. I am very grateful to Prof. Neil
Shafer--Ray for a fruitful discussions and University of Oklahoma for the
hospitality.

\newpage

\section*{Figure Captions}

\subsection*{Figure 1}

$\label{fi1}$Relative population distribution among magnetic substates$%
\left| J,M\right\rangle $ for angular momentum state with $J=20.$

\subsection*{Figure 2}

\label{fi2}Wave function evolution in an external magnetic field. 1 -- $%
\omega _Lt=0,$ 2 -- $\omega _Lt=\pi /4$, 3 -- $\omega _Lt=\pi /2$, 4 -- $%
\omega _Lt=3/4\pi $, 5 -- $\omega _Lt=\pi $, 6 -- $\omega _Lt=5\pi /4$

\subsection*{Figure 3}

\label{fi3}Stark quantum beats. 1 -- $t=0,$ 2 -- $t=\frac 15t_{re}$, 3 -- $t=%
\frac 14t_{re}$, 4 -- $t=\frac 12t_{re}$, 5 -- $t=\frac 34t_{re}$, 6 -- $%
t=t_{re}$


\begin{references}
\bibitem{av1}  Averbukh I. Sh., Perel'man N.F., ''The dynamics of wave
packets of highly-excited states of atoms and molecules'', Sov. Phys. Usp.
34 (7), 572--591 (1991)

\bibitem{al1}  Alberg G. and Zoller P., ''Laser Excitation of Electronic
wave packets in Rydberg atoms'', Phys. Rep. 199, 231--280 (1991)

\bibitem{ga1}  Garraway B.M. and Suominen K-A, ''Wave-packet dynamics: new
physics and chemistry in femto-time'', Rep. Prog. Phys. 58, 365--419 (1995)

\bibitem{au4}  Auzinsh Marcis, ''Angular momenta dynamics in magnetic and
electric field: Classical and quantum approach'', Can. J. Phys. 75, 853--872
(1997)

\bibitem{au1}  Auzinsh M. and Ferber R., Optical polarization of Molecules
(Cambridge University Press, Cambridge, U.K. 1995), p. 306

\bibitem{alx1}  Aleksandrov E.B., ''Quantum beats of luminescence under
modulated light excitation'', Opt. Spectrosc. (USSR), 14, 233--234 (1963)

\bibitem{do1}  Dodd J.N., Kaul R.D. and Warington D.M., ''The modulation of
resonance fluorescence excited by pulsed light'', Proc. Phys. Soc. London,
Sect. A, 84, 176--178 (1964)

\bibitem{va1}  Varshalovich D.A., Moskalev A.N., Khersonskii V.K., Quantum
Theory of Angular Momentum (World Scientific, Singapore, 1988), p. 514.

\bibitem{bl1}  Bluhm Robert, Kostelecky V.Alan, Porter James A, ''The
evolution and revival structure of localized quantum wave packets'', Am. J.
Phys. 64 (7), 944--953 (1996)

\bibitem{qb1}  Hack E. and Huber J.R., ''Quantum beat specrtoscopy of
molecules'', International Reviews in Physical Chemistry, 10, 287--317 (1991)

\bibitem{hi1}  Hilborn R.C., ''Atoms in orthogonal electric and magnetic
fields: A cpmparison of quantum and classical models'', Am. J. Phys. 63,
330--338 (1995)

\bibitem{au2}  Auzinsh M., Ferber R. and Stolyarov A., ''Separation of
quadratic and linear external field effects in high $J$ quantum beats'', J.
Chem. Phys. 101 (7), 5559--5565 (1994)

\bibitem{au3}  Auzinsh M.P., Ferber R.S., ''$J$-selective Stark orientation
of molecular rotation in a beam'', Phys. Rev. Lett., 69, 3463--3466 (1992)

\bibitem{fa1}  Fano U. and Racah G., Irreducible tensorial sets (Academic
Press, New York, 1959), p. 169.
\end{references}
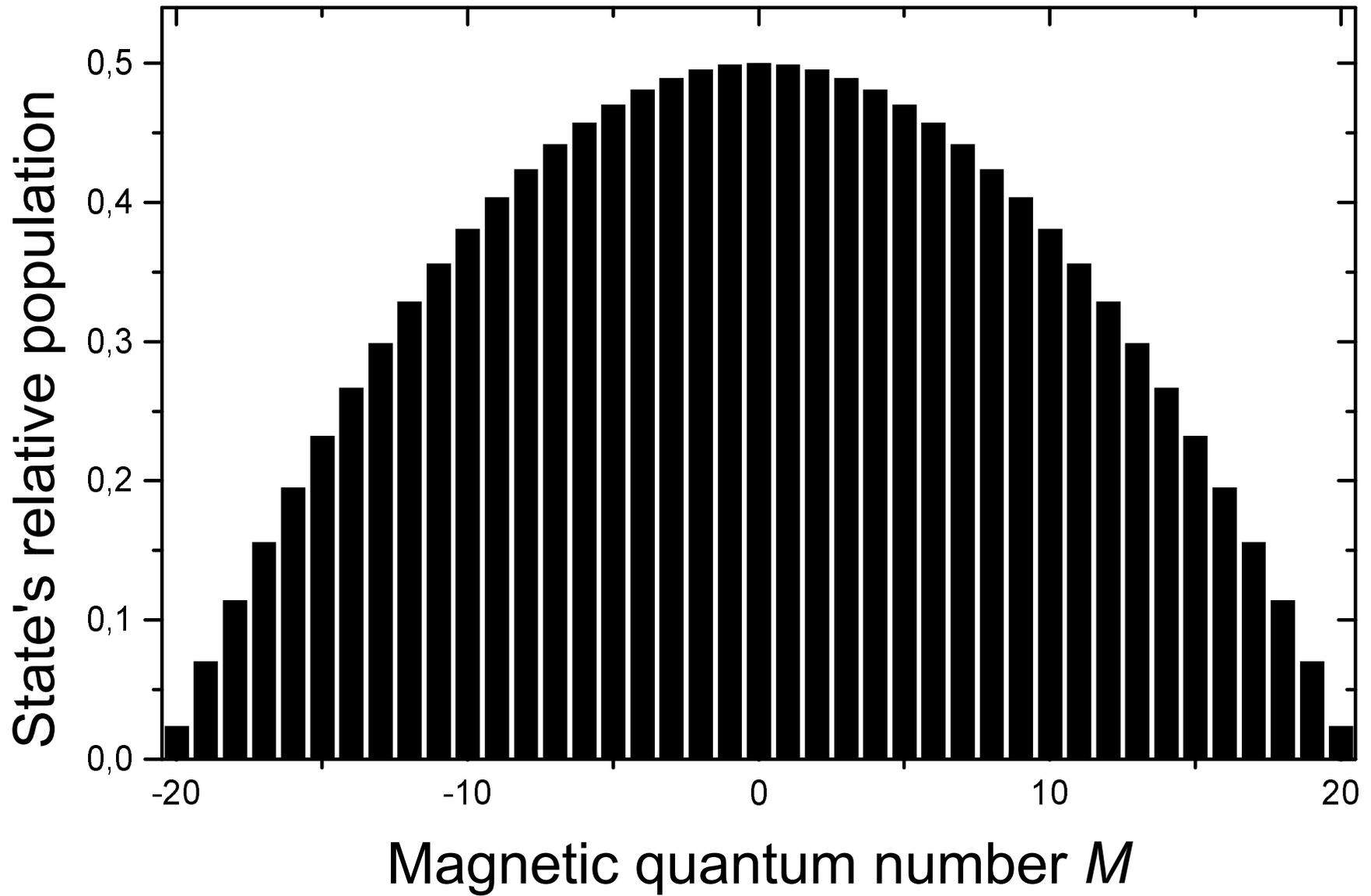
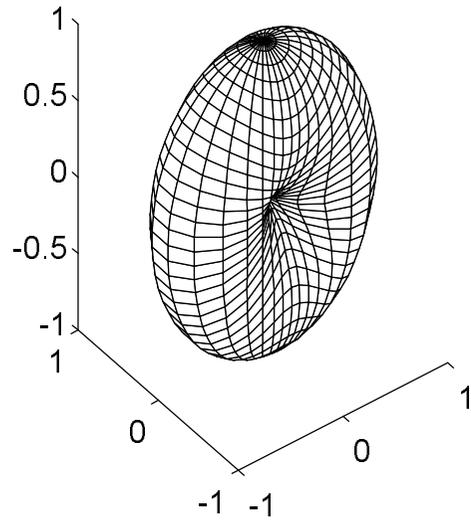
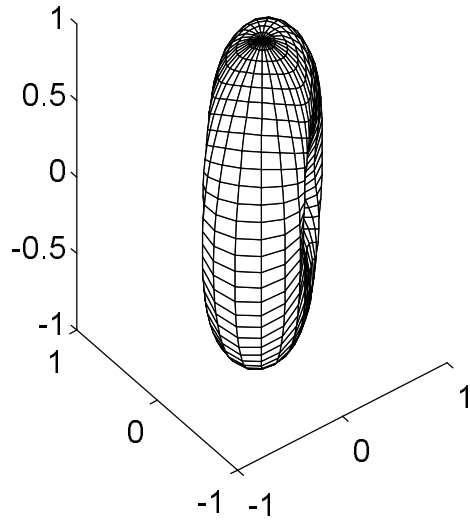
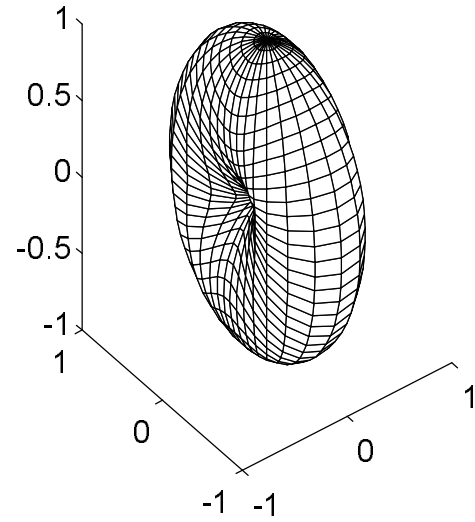
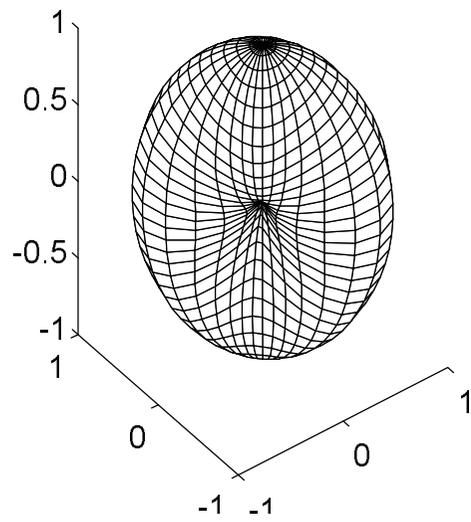
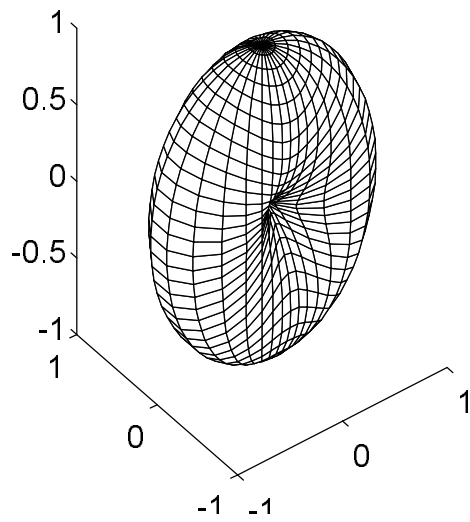
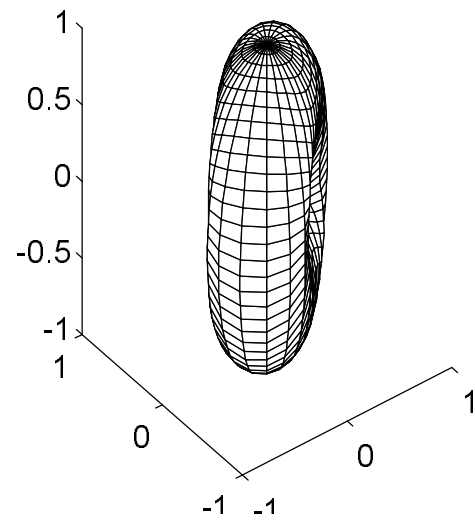
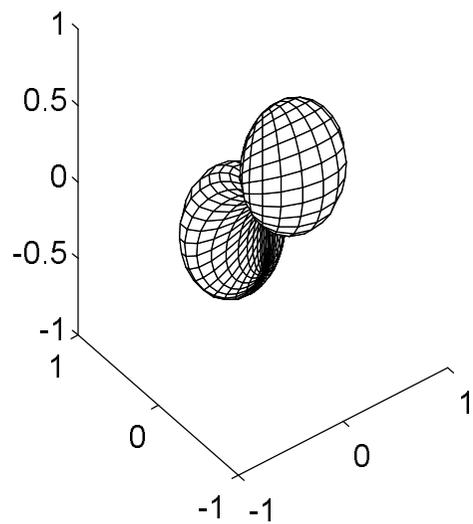
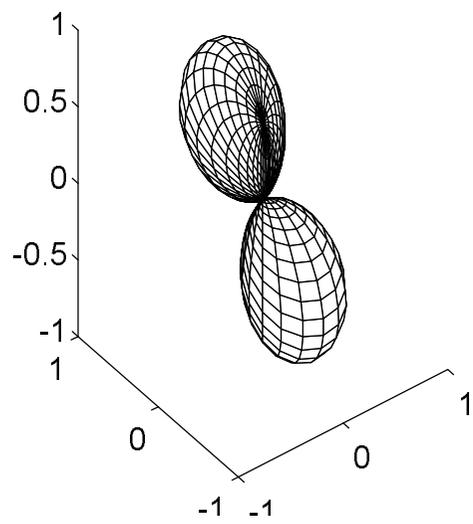
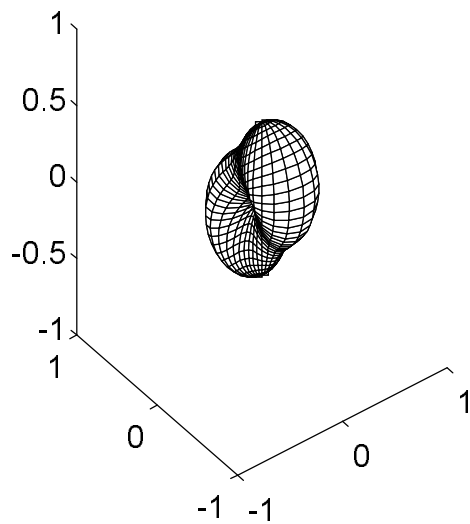
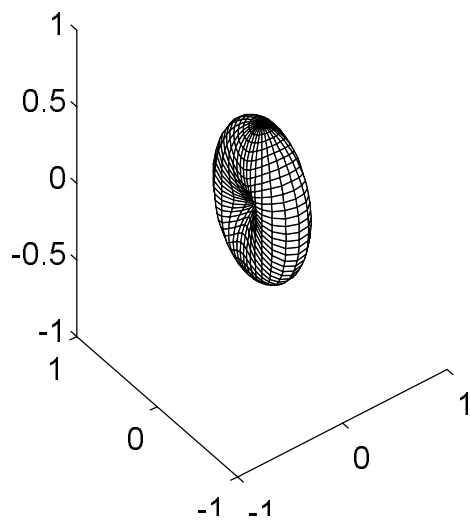
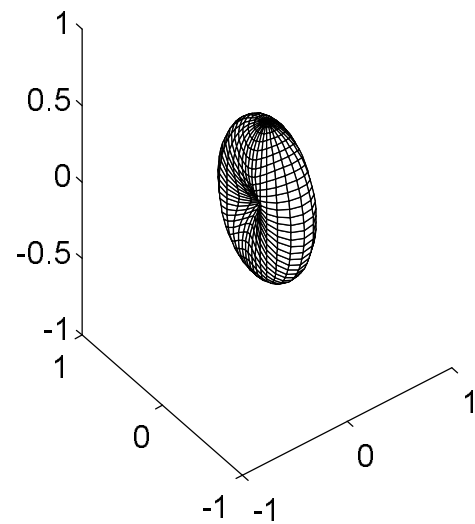
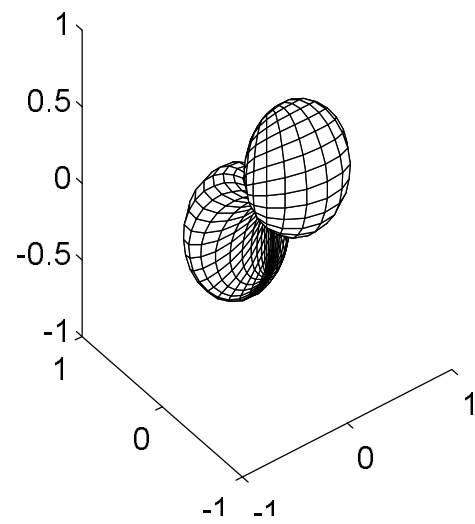
\end{document}